\documentclass[eqsecnum,amsmath,ams,showpacs]{revtex4}

\begin{document}

\title{Renormalization Group Approach to  Generalized Cosmological models}

\author{{J. Ib\'a\~nez $^1$, S. Jhingan$^{2,3}$}}

\affiliation{$^1$ Dpto. de F\'{\i}sica Te\'orica, Universidad del
Pa\'{\i}s Vasco, Apartado 644, 48080, Bilbao, Spain }

\affiliation{$^2$ Centre for Theoretical Physics, Jamia Millia
Islamia, Jamia Nagar, Delhi-110092, India}

\affiliation{$^3$ Department of Physics, Jamia Millia Islamia, Jamia
Nagar, Delhi-110092, India}

\date{{\small \today}}

\begin{abstract}
  We revisit here the problem of generalized cosmology using
  renormalization group approach. A complete analysis of these
  cosmologies, where specific models appear as asymptotic fixed-points,
  is given here along with their linearized stability analysis.
\end{abstract}

\pacs{98.80.Jk, 64.60.Ak}

\maketitle

\section{Introduction}

It is widely accepted that Einstein's general theory of relativity
is an effective four-dimensional theory at lower energies and that
it needs modifications at high energies. In the cosmological
context, this break down of the theory results in a modification of
the Friedmann equation with important consequences in the dynamical
evolution at early times (see for example, \cite{Martens, coley}).
There has been several proposals for cosmological scenarios mostly
inspired by brane-world models or by the presence of dark energy.
All of them lead to modifications to the total energy density
dependence of the Friedmann equation. Also, there have been studies
using dynamical system approach, bringing together several of such
corrections to the Friedmann equation, in order to study the
conditions which lead to attractor scaling solutions \cite{CLLM}.

In the dark energy scenarios, the energy density of the scalar
field, responsible for the dark energy, is important only at late
times and can account for the acceleration of the universe
\cite{Ratra_Peebles, Steinhardt, trac, CLW, ferreira, Liddle,
moduli_stabilization01, moduli_stabilization02,
moduli_stabilization03, sahni, delaMacorra:1999ff, Ng:2001hs,
Copeland:2000hn, Huey:2001ae, Sahni:2001qp, Huey:2001ah,
Majumdar:2001mm, Nunes:2002wz, vandenhoogen, Lidsey:2003sj}.
Therefore, scaling solutions (i.e. solutions for which the energy
density of the scalar field and the perfect fluid scale in the same
way), which appear as late time attractors of the system of the
evolution equations can  play an important role in modeling dark
energy scenarios \cite{Tsujikawa:2003zd}. The existence and behavior
of these models in the scaling regime depends on the proposed
modifications to the Friedmann equation, as well as on the form of
the scalar field potential.

In this paper we analyze asymptotic evolution of modified
cosmologies with a scalar field and a barotropic perfect fluid. The
modification of the standard cosmology is parameterized by a
function of the total energy density, in the line of Copeland et al.
\cite{CLLM}. The analysis of the asymptotic behavior of the
equations is performed using the techniques of the renormalization
group (RG), which has emerged as a powerful method for doing global
and asymptotic analysis of ordinary and partial differential
equations (Illinois group \cite{Illinois}, Bricmont et al
\cite{Bricmont}, Caginalp \cite{Caginalp}, Moise and Ziane
\cite{Moise}, for a reformulation of the RG approach in terms of the
classical theory of envelops see Kunihiro, \cite{Kunihiro}, the
theory of perturbations of an isotropic universe with dynamically
evolving Newton constant and Cosmologial constant \cite{Bonnano,
shapiro}, and \cite{koike, JJ} for application of RG techniques to
General Relativity).  We shall show in this paper that the RG
technique provides a very effective method for obtaining and
analyzing such solutions.

The plan of the paper is the following: In section II we describe
the basic equations governing generalized cosmology. This is
followed by a section where we give a very brief introduction to the
RG method and apply this technique to the generalized cosmological
models. Next section deals with a study of the exact scaling
solutions, which are the fixed-points of the RG transformations. The
stability of these exact solutions under linearized perturbations is
analyzed in section V. We conclude with a brief summary.

%----------------------------------------------------------------

\section{Generalized Cosmology}

Extending their earlier analysis of FLRW models using dynamical
system approach \cite{CLW}, Copeland et al.  showed that the
modified cosmological models, such as Randall-Sundrum,
Shatnov-Shani, etc.  can be recovered as scaling solutions by
modifying the Friedmann equation \cite{CLLM},
\begin{equation}
H^{2}=\frac{8\pi}{3m_{4}^{2}}\rho\,G^{2}\left(  \rho\right)  \,.
\label{mod-friedmann}
\end{equation}
For $G(\rho)= \sf{constant}$, we recover the standard FLRW
cosmology. Here $\rho$ is total energy density, and $m_4$ is the
four-dimensional Planck mass.

The model of universe considered here contains a barotropic fluid
$p_{\gamma}=(\gamma-1) \rho_{\gamma }$, with adiabatic index $0 \leq
\gamma \leq 2$, and a scalar field with energy density
\begin{equation}\label{scalar-density}
  \rho_{\phi} = \frac{1}{2}{\dot \phi}^2 + V( \phi ) ,
\end{equation}
where $V( \phi )$ is the scalar field potential and
$\rho_\gamma + \rho_\phi = \rho
$. From the Einstein equations we get the following system of
equations
\begin{equation}
\begin{tabular}[c]{lll}
  $ \dot{H} $ & $ = $ & $ -\frac{4\pi G}{m_{4}^{2}} \left(
    G+2\rho\frac{dG}{d\rho}\right) \left[
    \left(1-\frac{\gamma}{2}\right){\dot \phi}^2 + \gamma(\rho-V)\right] $ \\
  $ \dot{\rho} $ & $ = $ & $-3 H \left[
    \left(1-\frac{\gamma}{2}\right){\dot \phi}^2 + \gamma(\rho-V)\right] $ \\
  $ \dot \rho_\gamma  $ & $ = $ & $ - 3\gamma H \rho_\gamma  $ \\
  $ \ddot \phi $ & $ = $ & $ - 3 H \dot \phi -  \frac{dV}{d\phi} $ \; .
\end{tabular}
 \label{new-system}
\end{equation}
The evolution of scalar potential $ V $ and the generalization
function $ G $ depends on their respective functional dependence on
$ \phi $, and $\rho$ ,
\begin{equation*}
\dot V = \left( \frac{dV}{d \phi} \right) {\dot \phi} , \quad \dot G=
\left( \frac{dG}{d \rho}\right) \dot \rho \; ,
\end{equation*}
and which are for the moment left arbitrary.  The generalization function
$ G $ modifies the Hubble parameter, changing the cosmological
expansion/contraction rate ($H = {\dot a}/a$), and that
correspondingly alters the evolution of all the physical variables
in the system.

Before applying the RG procedure to the above set of evolution
equations some comments on scaling solutions are in order. It is
well known that the scalar fields with exponential potential admit
scaling solutions. The form of the potential changes in the case of
generalized cosmologies; Copeland et al. in ref. \cite{CLLM} gave a
necessary and sufficient condition relating the scalar field
potential and the generalization function $ G $ in order to get
scaling solutions. Instead of repeating their procedure to get the
required condition, we give below an alternative derivation to
obtain an equivalent condition.

Let's  introduce a new variable
\begin{equation*}
x = a^{-3 \gamma} .
\end{equation*}
Essentially we want to use the scale factor $ a(t) $, which is related
to Hubble parameter, instead of the usual coordinate time $t$, as our
evolution parameter.  In the equation set (\ref{new-system}), the
equation giving evolution of the energy density of the fluid can be
integrated trivially in the new parametrization as
\begin{equation}\label{in-grho}
  \rho_{\gamma} = \rho_{o\gamma} x
\end{equation}
where $ \rho_{o\gamma} $ is an integration constant. The argument
leading to the desired relation between $V$ and $G$ is now
straight forward. Restricting ourselves to scaling solutions, i.e.
both the scalar field density and the fluid density scale in the
same way, we have
\begin{equation}\label{scal-cond}
  \rho_{\phi} \sim \rho_{\gamma} \quad \Rightarrow \quad \rho_{\phi}
  = \rho_{o\phi}x  \quad \Rightarrow \quad \rho = \rho_{o} x
\end{equation}
where $ \rho_{o} = \rho_{o\phi} + \rho_{o\gamma} $. Using the
evolution equation of the total density along with the scaling
condition above, the form of potential can now be determined as
\begin{equation}\label{form-pot}
  V = \frac{(2-\gamma)}{2} \rho_{o\phi} x.
\end{equation}
Also, using definition of $ \rho_{\phi} $, the kinetic energy term
for the scalar field can be recovered as,
\begin{equation}\label{form-KE}
 {\dot \phi}^2 = \gamma \rho_{o\phi} x.
\end{equation}
On the other hand, we have
\begin{equation}
 {\dot \phi}=-3\gamma xH\frac{d\phi}{dx} .
\end{equation}
Taking the square of the above equation, substituting
(\ref{mod-friedmann}) and (\ref{form-KE}), and  bearing in mind
that now function $ G $ depends on $ x $, we get
\begin{equation}
  3\sqrt{\frac{8\pi}{3 {m_4}^2}} \sqrt{\frac{\gamma
      \rho_{o}}{\rho_{o\phi}}} \quad \phi = \pm \int\frac{dx}{xG(x)} .
\end{equation}
Once the function $ G $ is fixed, the above equation gives $ \phi
$ as a function of $ x $. Now taking the inverse of this function
and substituting it in (\ref{form-pot}) the form of potential
corresponding to the specific choice of $G$ is obtained. The
condition given above is equivalent to the one obtained by
Copeland et al. \cite{CLLM}. Therefore, condition for the scaling
solution, relating generalization parameter $G$ with potential
$V$, can be easily recovered starting with the scaling ansatz,
that is, both the scalar field component and the fluid component
scale in the same way.

\section{Renormalization Group Approach}

The renormalization group (RG) approach developed by Wilson (see for
example, Wilson and Kogut \cite{wilson-kogut}) has enjoyed
tremendous success, and is seen as a broad philosophy rather than
just a mathematical technique. In the past decade this philosophy is
also applied with considerable success to understanding some very
basic aspects of non-linear differential equations. In particular,
equivalence of RG and intermediate asymptotics was shown by
Goldenfeld et. al. (see for example, \cite{Illinois}). They showed
that the so called anomalous dimension in RG theory are actually the
non-trivial exponents appearing in the intermediate asymptotics.

The scaling solutions which we are concerned with are exact
solutions, and which exhibit self-similarity at an asymptotic
fixed-point.  Therefore, RG seems to be a natural choice for our
purpose.  The form in which RG is used here was developed by
Bricmont and Kupiainen \cite{Bricmont}. The basic idea is as
follows: We do a finite time integration followed by a rescaling of
variables in the problem. This scaling transformation together with
evolution equations gives us the RG equations. And, the fixed-points
of these equations are actually the scale-invariant solutions to
differential (evolution) equations which we are interested in.

We will apply now the RG method to the system of equations
governing the generalized cosmology, given in the previous
section.  Our interest here is in the early or late time regime,
where solution asymptotes to the form
\begin{equation}
\lim_{t\to \infty} u(t) = t^{\chi}u^{\ast}(1) \label{asymptotic-u} .
\end{equation}
Here $ u(t) $, refers to any variable of the former system of
equations, namely $H, \rho, \phi, V, G$. The value ``one'' in the
argument of $ u^{\ast}(1) $, signifies the initial value of the $ u
$, and $ \chi $ is the scaling exponent to be determined later. It
is convenient to choose $ t=1 $ as initial time.  The RG method
gives us a systematic procedure to determine the exponent $ \chi $,
and $ u^{\ast}(1) $, as we have illustrated below for the case of
generalized cosmological models. Apart from the large time decay,
this RG procedure is also applicable to the finite time decay or
blow up of solutions, where self-similarity is exhibited at the
asymptotic fixed-point.

%%%%%%%%%%%%%%%%%%%%%%%%%%%%%%%%%%

Let us consider following scale transformations
\begin{eqnarray}\label{gen-scalings}
  t \to   L t, \quad H \to  L^{e}H(Lt) \equiv H_L, \quad
  \rho  \to  L^{a}\rho(Lt) \equiv \rho_{L}\nonumber \\
  {\dot \phi} \to  L^{b}{\dot \phi} (Lt) \equiv {\dot \phi}_{L}, \quad
  V\to  L^{c}V \equiv V_L, \quad G \to    L^{d}G \equiv G_L
\end{eqnarray}
We use here a number $ L > 1 $, as a parameter for scale
transformations, and the quantities subscripted with index ``$L$'' are
scaled quantities.  The RG transformation ${\cal R}_L$ is defined as a map
from one initial data set to another \cite{koike},
\[
{\cal R }_L u(x,1) = u_L(x,1)  .
\]
These transformations have the semi-group property $ {\cal R}_{L^n}
= {\cal R}_{L^{n-1}} \circ {\cal R}_{L} $. Since the scaled
quantities satisfy the original system of equations, this fixes the
following exponents
\begin{equation}
  e=1,\, a= 2b = c,\, d = 1-\frac{a}{2} \; . \label{gen-scal-exp}
\end{equation}
Moreover, letting $t=1$ and $L = t$, now we can express solution at
arbitrary time in terms of initial values
\begin{equation}\label{g-sc-beh}
  H(t) = t^{-1} H_L(1), \; \rho(t)  = t^{-a} \rho_L(1), \; {\dot
\phi}(t) =
  t^{-\frac{a}{2}} {\dot \phi_L}(1), \; V(t) = t^{-a} V_L,\;  G(t) =
 t^{\frac{a}{2}-1} G_L
\end{equation}
where quantities sub-scripted with $ L $ are constants. The exponent $
a $ is the anomalous dimension: the field equations alone do not fix
it and it should be determined by initial or boundary conditions. For
instance, when one chooses a particular model by fixing the function $
G $, parameter $ a $ gets determined.

Note from Eq. (\ref{g-sc-beh}) that the evolution of $\phi$ (as well
as $\dot \phi$) depends on the value of parameter $ a $. This
behavior, as we will see later, has consequences on the nature of
the potential in the scaling regime. In particular it is the
parameter $a$, which distinguishes FRW standard cosmology from other
``generalized'' models.

Comments are now in order about nature of the potential $V$.  Though
the functional form of the potential is left free, clearly in the
scaling regime it gets fixed, Eq. (\ref{g-sc-beh}).  Moreover, the
functional form of $G$ also is not arbitrary. As argued in the
previous paragraph, depending on the value of parameter $a$, the
scaling regime can be divided in two categories; $a = 2$, which
gives the usual FLRW models and $ a \neq 2 $, the generalized
cosmology.

\noindent \underline{${\bf a=2}$} : From Eq.  (\ref{g-sc-beh}), we get
\begin{equation}\label{FLRW-fp}
  V(\phi) = V_0\exp (-2 {\phi}/{\dot \phi_L} ) , \quad
  G(\rho) = G_0 .
\end{equation}
Where $ V_0 = V_L(1) $ and $ G_0 = G_L(1) $ are constants. Therefore,
in the scaling regime we have a exponential potential and the
generalization function $G$ is a constant, i.e., FLRW cosmology.

\noindent \underline{${\bf a\neq 2} $} : In this case $G$ and $V$ have a
power law dependence on $\rho$ and $\phi$, respectively, in the scaling
regime
\begin{equation}\label{GFLRW-fp}
  V(\phi) = V_0 {\phi_L}^{\frac{2a}{a-2}} , \quad G(\rho) =
  G_0 \rho_L^{\frac{2-a}{2a}} ,
\end{equation}
where $G_0$ and $V_0$ have simple dependence on the $L$ sub-scripted
quantities.

We draw a comparison now between the scaling solutions in cosmology,
discussed in the previous section, and the scale invariant solutions
which arise as asymptotic fixed-points of the RG analysis.  First
note that we are using as a variable the total energy density
$\rho$, instead of the density of the perfect fluid $\rho_{\gamma}$.
However, from Eq. (\ref{new-system}) we have the following scaling
relation for the perfect fluid energy density $\rho_{\gamma} \sim
t^{-a}$. It is easy now to see that the energy density of the
perfect fluid and the energy density of the scalar field scale in
the same way
\begin{equation}
  \frac{\rho_\phi}{\rho_{\gamma}} = \frac{\dot {\phi_L}^2 +
    2 V_L}{ 2 {\rho_{\gamma}}_L} = \sf{constant}
\end{equation}
which is the usual definition of scaling solutions in scalar
field cosmologies. Therefore, in RG method we recover all the scaling
solutions of the system as a subset of scale invariant solutions,
provided that both components of energy density are non-vanishing.

The large $L$ in our notation means late time. Therefore, applying the
RG transformation repeatedly we can recover the long time behavior of
the solution. Defining a auxiliary parameter $\tau$ through
$L=\exp(\tau)$, the RG equations are
\[
 \frac{dH_L}{d\tau} = H_L + \left.\frac{dH_L}{dt} \right|_{t=1} =
H_L - \left\{\frac{4\pi G_L}{m_{4}^{2}} \left(
    G_L+2\rho_L\frac{dG_L}{d\rho_L}\right) \left[
    \left(1-\frac{\gamma}{2}\right){\dot \phi}_L^2 + \gamma(\rho_L-V_L)\right]
\right\}_{t=1}
\]
\[
 \frac{d\rho_L}{d\tau} = a \rho_L + \left. \frac{d\rho_L}{dt}
\right|_{t=1} = a \rho_L - \left\{3H_L \left[
\left(1-\frac{\gamma}{2}\right){\dot \phi}_L^2
    + \gamma(\rho_L - V_L)\right] \right\}_{t=1}
\]
\begin{equation}\label{gen-rg-rqns}
  \frac{d{\dot \phi}_L}{d\tau} = \frac{a}{2} {\dot \phi}_L +
  \left.\frac{d{\dot \phi}_L}{dt} \right|_{t=1} =
  \frac{a}{2}{\dot \phi_L} - \left\{ 3 H_L {\dot \phi_L} -
    \frac{dV_L}{d\phi_L}\right\}_{t=1}
\end{equation}
\[
 \frac{dV_L}{d\tau} = a V_L + \left.\frac{dV_L}{dt} \right|_{t=1}
= a V_L + \left\{\left(\frac{dV_L}{d\phi_L}\right) {\dot
    \phi_L}\right\}_{t=1}
\]
\[
 \frac{dG_L}{d\tau} = \left(1-\frac{a}{2}\right)G_L +
\left.\frac{dG_L}{dt} \right|_{t=1}= \left(1-\frac{a}{2}\right)G_L +
\left\{ \left(\frac{dG_L}{d\rho_L}\right){\dot \rho_L}\right\}_{t=1}
.
\]
Which is a set of algebraic equations since the quantities on the
right hand side are all evaluated at $t=1$. Note that the system of
equations above has no dependence on $L$; this is expected since the
system (\ref{new-system}) is scale invariant.

\section{Scale invariant solutions}

The scale invariant solutions appear as fixed-points to the RG
equations. The fixed-points are those points which are mapped onto
themselves by the RG transformations for any $ L>1 $
\[
{\cal R}_{L} u^{\ast} = u^{\ast} ,
\]
and which is equivalent to
\[
\frac{du^{\ast}}{d\tau} = 0.
\]
Therefore, the equations determining the fixed-points of our system
are
\begin{equation}
  \frac{dH_L^{\ast}}{d\tau}=0,\,\quad
  \frac{d\rho_L^{\ast}}{d\tau}  =0,\quad
  \frac{d\dot{\phi}_L^{\ast}}{d\tau}=0,\quad \frac{dV_L^{\ast}}{d\tau}=0 .
\label{fix-point-eqns}
\end{equation}
As discussed earlier in this paper, if classified in terms of the
potential function of the scalar field there are two classes of
solutions corresponding to the exponential and the power-law
potentials; equivalently the system can also be classified in terms
of the generalization function $G$. Moreover, these two different
classes also differ in their value of anomalous dimension ``$ a $''
and which, therefore, can also serve as an alternative way of
classifying these fixed-points.

\subsection{Exponential Potential ($a = 2$)}

To facilitate a comparison with earlier works in FLRW scalar field
cosmology we shall use $\rho_{\gamma}$, as a variable in this
subsection instead of the total energy density $\rho$. Moreover, we
work with the notation used in \cite{CLW}, defining $\kappa = 8\pi
G$.  The set of RG equations simplifies to
\[
  \frac{dH_L}{d\tau}  =
  H_L-\left\{\frac{\kappa^{2}}{2}(\gamma\rho_{\gamma L} +
  {\dot{\phi_L}^{2}})\right\}_{t=1}
\]
\begin{equation}
  \frac{d\rho_{\gamma L}}{d\tau}  =  2 \rho_{\gamma L} - 3\gamma \left\{ \rho_{\gamma L}
  H_L\right\}_{t=1} \\
\label{rg-equations}
\end{equation}
\[
  \frac{d{\dot{\phi_L}}}{d\tau}  =
  {\dot{\phi_L}} - 3  \left\{{\dot{\phi_L}} H_L - \frac{dV_L}{d\phi_L} \right\}_{t=1}
\]
Therefore, the fixed-points for this system are
\begin{equation}
  \frac{dH_L^{\ast}}{d\tau}=0,\,\quad
  \frac{d\rho_{\gamma L}^{\ast}}{d\tau}
  =0,\quad\frac{d\dot{\phi_L}^{\ast}}{d\tau}=0 .
\label{fix-point2}
\end{equation}
The fixed-points should also satisfy the Friedmann equation
\begin{equation}
  {H^\ast}^2=\frac{\kappa^2}{3}(\rho_{\gamma}^{\ast}+\frac{1}{2}{\dot {\phi^{\ast}}}^2+V^{\ast}) .
\label{friedmann}
\end{equation}

{}From the second equation in Eq. set (\ref{fix-point2}), we see that
there are two different sets of fixed-points: those with $
\rho_{\gamma}^{\ast}=0 $ and those with $ \rho_{\gamma}^{\ast} \neq 0 $.  Let's start
with the fixed-points characterized by $ \rho_{\gamma}^{\ast} = 0 $. This case
can be further subdivided into two sub-cases. In the first case,
corresponding to $V^\ast=0$, we obtain
\begin{equation}
  V^\ast=0, \quad \left(\frac{dV}{d\phi}\right)^\ast=0, \quad
H^\ast=\frac{1}{3},\quad \rho_{\gamma}^{\ast}=0, \quad {\dot
\phi}^\ast= \pm\frac{1}{\kappa}\sqrt{\frac{2}{3}} .
\end{equation}
This solution corresponds to a massless scalar field cosmology.

In the second case, $V^{\ast} \neq 0$, the nature of potential in the
neighborhood of the fixed-point is given by (\ref{FLRW-fp}).
Comparing the potential (\ref{FLRW-fp}) with the exponential potential
used in \cite{CLW}, $V=V_0 \exp(-\kappa\lambda\phi)$, the fixed-point is given
by
\begin{equation}\label{fixed-2}
  {\dot \phi}^\ast=\frac{2}{\kappa\lambda},\quad \rho_{\gamma}^{\ast}=0,\quad
  H^\ast=\frac{2}{9 \lambda^2},\quad V^\ast=
  \frac{2}{\kappa^2 \lambda^2}\left(\frac{6}{\lambda^2}-1\right) .
\end{equation}
It is a exponential potential scalar field cosmology.

For the fixed-points with non-zero perfect fluid $ \rho_{\gamma}^\ast \neq 0 $
the value of Hubble parameter is determined $ H^\ast = 2/3\gamma $. And we
have following sub-cases due the nature of the potential. One case
with $V^\ast=0$, and the solution for fixed-point is
\begin{equation}
  H^{\ast} =  \frac{2}{3\gamma}, \quad \rho_\gamma^\ast=\frac{4}{3\kappa^2\gamma^2},\quad
  \dot\phi^\ast=0, \quad V^{\ast} = 0 .\label{fixed-3}
\end{equation}
Which is a perfect fluid only cosmology. The second case $V^\ast
\neq 0$,
 the fixed-point is
\begin{equation}
  H^{\ast} = \frac{2}{3\gamma}, \quad  \rho_\gamma^\ast=\frac{4}{\kappa^2\gamma^2}
  \frac{\gamma - 3\lambda^2}{9\lambda^2}, \quad
  {\dot \phi}^\ast=\frac{2}{3\kappa\lambda},\quad
  V^\ast = \frac{2(2-\gamma)} {9\kappa^2\gamma\lambda^2},\label{fixed-4}
\end{equation}
for $ \lambda^2 > 3\gamma $.

The fixed-points obtained are the same as in given in \cite{CLW},
where the scalar field potential is assumed to be of exponential
form from the very beginning, and the system of equations was
written in terms of normalized variables that make the phase space
of the autonomous system bounded. It is interesting to note that
imposing scale invariance we obtain the same results without using
normalized variables. On the other side, it is reasonable to think
that, what are called scaling solutions should be scale invariance
solutions, since in that case the scalar field energy density and
the energy density of the matter scale as a power of the scale
factor. Assuming this we have proved that the only potential that
gives scale invariance solutions is of exponential type in FLRW
cosmology.

\subsection{Power-law potentials ($ a \neq 2$)}

In deriving the expression for potential in the scaling regime we have
used the scaling behavior of $\phi$, i.e.,
\[
\frac{d\phi_L}{d\tau} = \left(\frac{a}{2}-1\right)\phi_L + {\dot
\phi_L}
\]
implying $2{\dot\phi}^{\ast} = (2-a){\phi}^{\ast} $ at the fixed-point.

Let's note also that, as in the standard cosmology, there are two
classes of fixed-points depending on whether $\rho_\gamma$ vanishes or not.
The quantity $\rho_\gamma$, however, is not a variable in our system of
equations (\ref{gen-rg-rqns}) (we have used $\rho$ instead),
nevertheless $\rho_\gamma$ satisfies the following equation
\[
\frac{d\rho_\gamma}{d\tau}=\rho_\gamma(a-3\gamma H) ,
\]
which means that if fluid density is non-vanishing at the fixed point
$\rho^{\ast}_\gamma\neq 0$, the Hubble parameter takes the critical value
$H^{\ast}=a/3\gamma$.

To analyze the fixed-points when $a\neq 2$, we first consider $V = 0$.
In this case the RG equations simplify considerably
\[
{H^{*}}^2  =  \frac{8\pi G_0^2}{3m_4^2}{\rho^{\ast}}^{\frac{2}{a}}
\]
\begin{equation}
    2 a\rho^{\ast}  =  3H^{\ast} [(2-\gamma) {\dot \phi}^{\ast}{}^2 +
    2\gamma\rho^{\ast}]
\end{equation}
\[
a {\dot \phi}^{\ast}  =  6 H^{\ast} {\dot \phi}^{\ast}   \\
\]
giving the following set of fixed-points
\begin{eqnarray}
{\dot \phi}^{\ast} = 0, \qquad \rho^{\ast} = \left(\frac{a^2
      m_4^2}{24\pi \gamma^2G_0^2}\right)^{\frac{a}{2}}, \qquad H^{\ast}
  = \frac{a}{3\gamma} \label{fl_only} \\
  \label{sc_only} {\dot \phi}^{\ast} = \pm \sqrt{2\rho^{\ast}},
  \qquad \rho^{\ast} = \left(\frac{a^2 m_4^2}{96\pi G_0^2}
  \right)^{\frac{a}{2}}, \qquad H^{\ast} = \frac{a}{6}
\end{eqnarray}
The fixed-point (\ref{fl_only}) corresponds to vanishing of the
scalar field and (\ref{sc_only}) to the vanishing of the perfect
fluid. Therefore, only one component survives in the scaling regime.
Note that (\ref{sc_only}) is not a limiting case of (\ref{fl_only})
for $\gamma = 2$.

The fixed-points in the more general case with non-zero potential
are governed by the following set of algebraic equations
\[
{H^{*}}^2  =  \frac{8\pi G_0^2}{3m_4^2}{\rho^{\ast}}^{\frac{2}{a}}
\]
\begin{equation}
8 a\rho^{\ast}  =  3H^{\ast} [(2-\gamma)(2-a)^2 {\phi^{\ast}}^2 +
    8\gamma(\rho^{\ast}-V_0 ~{\phi^{\ast}}^{\frac{2a}{a-2}})]
\end{equation}
\[
H^{\ast} \phi^{\ast} =  \frac{a}{6} \phi^{\ast}   -
    \frac{4a V_0}{3(a-2)^2} {\phi^{\ast}}^{\frac{a+2}{a-2}}
\]
These cannot be solved in general however various special cases
corresponding to different values of parameter $a (\neq 2)$ can be
analyzed.

First we consider the simplest scenario where potential $V$ vanishes
in the asymptotic regime due to vanishing of the scalar field itself
($ \phi^{\ast} = 0 $). This is unlike the previous case where potential was
identically zero (due to, say, $V_0 = 0$) to begin with. We recover
following fixed-point for vanishing of the scalar field $\phi^{\ast}=0$,
\begin{equation}
  H^{\ast} = \frac{a}{3\gamma},  \rho^{\ast} = \left(\frac{a^2
      {m_4}^2}{24\pi\gamma^2 G_0^2} \right)^{\frac{a}{2}},  V^{\ast} =0
\end{equation}
This is similar to the previous case, except the fact that the range
of anomalous dimension $a$ is restricted to $|a| > 2$, and that the
cosmology is driven towards vanishing potential which is non-zero to
begin with.

The fixed-points corresponding to a non-vanishing potential in the
scaling regime can be divided in two families. The one corresponding
to vanishing fluid component in scaling regime is given by
\begin{equation}
  {H^{*}} = \pm \sqrt{\frac{8\pi
      G_0^2}{3m_4^2}}{\rho^{\ast}}^{\frac{1}{a}}, \quad \rho^{\ast} =
  \left[\frac{3\pi (2-a)^4 G_0^2}{2a^2m_4^2}\right]^{\frac{a}{2(a-1)}}
  {\phi^{\ast}}^{\frac{2a}{a-1}}, \quad V^{\ast} = V_0
  {\phi^{\ast}}^{\frac{2a}{a-2}} .
\end{equation}
The plus and the minus signs correspond to expanding and
contracting models, respectively, and the $\phi^{\ast}$ in the
equation above is given by the solution to following algebraic
equation ($a \neq 1, 2$) ;
\begin{equation}
  {\phi^{\ast}}^{\frac{4}{a-2}}  - \frac{1}{V_0} \left( 1 -
    \frac{a}{2}\right)^{\frac{2a}{a-1}} \left[\pm \sqrt{\frac{24\pi
        G_0^2}{a^2m_4^2} }\right]^{\frac{a}{(a-1)}}
  {\phi^{\ast}}^{\frac{2}{a-1}} +\frac{(2-a)^2}{8 V_0} = 0 .
\end{equation}
The case $a=1$ needs to be treated separately and the fixed-point
is of the form
\begin{equation}\label{aeq1}
 {H^{*}} = \pm \sqrt{\frac{8\pi G_0^2}{3m_4^2}}\rho^{\ast}, \quad
\rho^{\ast} = \pm \sqrt{\frac{3\pi G_0^2}{2
    m_4^2}}\left(\frac{m_4^2}{12\pi G_0^2}+V_0\right), \quad
{\phi^{\ast}}^2 = \pm \sqrt{\frac{2 m_4^2}{3\pi G_0^2}}, \quad
V^{\ast} = V_0 {\phi^{\ast}}^{\frac{2a}{a-2}} \;.
\end{equation}
The other fix-point corresponding to non-vanishing of both the scalar
field and the perfect fluid is of the form
\begin{equation}\label{non_zero}
 H^{\ast} = \frac{a}{3\gamma} , \quad \phi^{\ast} = \left[
    \frac{(2-\gamma)(2-a)}{4\gamma V_0}\right]^{\frac{a-2}{4}}, \quad
  \rho^{\ast} = \left(\frac{a^2 {m_4}^2}{24\pi\gamma^2 G_0^2}
  \right)^{\frac{a}{2}},  \quad V^{\ast} = V_0
  {\phi^{\ast}}^{\frac{2a}{a-2}} .
\end{equation}

For $a=2$ system reduces to usual FLRW models with exponential
potentials which were analyzed in the previous subsection.

\subsection{Stability}
The scaling solutions studied in the previous section arise as
fixed-points of the RG equations. The stability in time evolution
around a self-similar solution is stability of RG flow near the
fixed-point. We do a linear perturbation analysis here, considering
perturbations in the initial data space staying close to the
fixed-points and searching for relevant modes, i.e., modes which make
the flow diverge from the fixed-points.

Let us define following perturbation quantities
\[
  H = H^{\ast}+\delta H, \; \rho = \rho^{\ast} + \delta \rho, \;
{\dot
    \phi} = {\dot \phi}^{\ast} + \delta {\dot \phi}, \; \phi =
  \phi^{\ast} + \delta \phi, \;V = V^{\ast} + \delta V , \; G =
  G^{\ast} + \delta G
\]
Where all perturbed quantities are small ($\ll 1$). The Friedmann
equation can be linearized as
\begin{equation}\label{pert_H}
  H^{\ast}\delta H = \frac{4\pi}{3m_4^2}G^{\ast}[ G^{\ast} \delta
  \rho+2{\rho^{\ast}}\delta G]
\end{equation}
and can be used to relate variation in $H$ with other variables.
However, this reduction in variables can make the equation set much
more complicated and for the moment we will work with the full set of
variables.

The linearized RG equations take the following form
\begin{eqnarray}
  \frac{d\delta \rho}{d\tau} & = & (a-3\gamma H^{\ast})\delta
\rho +
  3(\gamma-2)H^{\ast}{\dot \phi}^{\ast} \delta {\dot \phi} + 3
  \gamma H^{\ast} \delta V -\frac{3}{2}\left[(2-\gamma)({\dot
      \phi}^{\ast})^2+2\gamma(\rho^{\ast}-V^{\ast})\right] \delta H  \nonumber\\
  \frac{d\delta{\dot\phi}}{d\tau} & = & \left(\frac{a}{2}-3
    H^{\ast}\right)\delta {\dot\phi} -3{\dot\phi}^{\ast} \delta H +
  \frac{2a}{(a-2)}\frac{V^{\ast}}{({\dot\phi}^{\ast})^2} \delta \phi
  + \frac{2a}{(2-a){\dot\phi}^{\ast}}\delta V \nonumber\\
  \frac{d\delta\phi}{d\tau} & = & \left(\frac{a}{2}
    -1\right)\delta\phi +\delta {\dot \phi} \nonumber\\
  \frac{d\delta V}{d\tau} & = &
  \frac{2a}{(a-2)}\frac{V^{\ast}}{\phi^{\ast}} \delta{\dot\phi} -
  \frac{2a}{(a-2)}\frac{V^{\ast}{\dot\phi}^{\ast}}{(\phi^{\ast})^2}
  \delta{\phi} + a
  \left(1+\frac{2}{(a-2)}\frac{{\dot\phi}^{\ast}}{\phi^{\ast}}\right)
  \delta V \nonumber\\
  \frac{d\delta G}{d\tau} & = & \frac{(a-2)}{2a}\left\{ -\left[
(a-3\gamma
    H^{\ast}) + \frac{3}{2}
    \frac{H^{\ast}}{\rho^{\ast}}\left[(\gamma-2)({\dot\phi}^{\ast})^2+2\gamma
      V^{\ast}\right] \right] \delta G + \frac{3}{2}
  \frac{G^{\ast}H^{\ast}}{(\rho^{\ast})^2}
  \left[(\gamma-2)({\dot\phi}^{\ast})^2 \right. \right.\nonumber \\
  & & \left.+ 2\gamma V^{\ast} \right]
  \delta \rho + \left.3 \left[\gamma G^{\ast} - \frac{1}{2}
    \frac{G^{\ast}}{\rho^{\ast}}\left[(\gamma-2)({\dot\phi}^{\ast})^2+2\gamma
      V^{\ast}\right] \right] \delta H +
  {3(2-\gamma)}\frac{H^{\ast}G^{\ast}{\dot\phi}^{\ast}}{\rho^{\ast}}
  \delta {\dot\phi} \right\}\nonumber .
\end{eqnarray}
We now find the normal modes with following ansatz
\begin{equation}
  \delta f = {\bar f} \exp(\omega\tau)
\end{equation}
where $f$ is a representative variable for $H, \rho, {\dot\phi}, \phi, V$
and $G$, and over bar signifies quantity to be constant.  With this
ansatz and also simplifying system using (\ref{pert_H}) we can now
look for any relevant modes.

Of all the fixed-points listed in the previous section the one given
by (\ref{non_zero}) is of particular interest since this corresponds
to late time regime where both scalar field as well as perfect fluid
component are non-vanishing. Taking these particular fixed-point
values the stability of the system depends of the following matrix
\\

\[
 \left(
\begin {array}{c|c|c|c|c}
  \omega + \frac{12 \pi {\gamma}^{2}}{a\, m_4^2} {G^{\ast}}^{2} \rho^{\ast} &
  \frac{a(\gamma-2)( a-2)}{2 \gamma} \phi^{\ast} & -a & 0 & \frac{24 \pi \gamma^2}{a
    m_4^2} G^{\ast}{\rho^{\ast}}^{2} \\ \hline \frac{6\pi\gamma(2-a)}{a m_4^2}
  {G^{\ast}}^2 \phi^{\ast} & \omega + \frac{a(2-\gamma)}{2\gamma} & \frac{a
    (a-2)(\gamma-2)}{4\gamma} & \frac{2 a}{(a - 2) \phi^{\ast}} & \frac{12 \pi \gamma
    (2-a)}{a m_4^2} G^{\ast} \rho^{\ast}
  \phi^{\ast} \\ \hline 0 & -1 & \omega +1 - \frac{a}{2}& 0 & 0 \\
  \hline 0 & \frac{a (a-2) (\gamma-2)}{4 \gamma} \phi^{\ast} &
  \frac{a (a-2)^2 (\gamma-2)}{8\gamma} \phi^{\ast} & \omega & 0 \\
  \hline \frac{6\pi \gamma^2 (2-a)}{a^2 m_4^2} {G^{\ast}}^3 &
  \frac{(2-\gamma)(a - 2 )^{2}}{4\gamma} \frac{ G^{\ast}\phi^{\ast}}{\rho^{\ast}} & 0
  & \frac{(a - 2)}{2} \frac{G^{\ast}}{\rho^{\ast}} & \omega - \frac{12 \pi
    \gamma^2 (a-2)}{a^2 m_4^2} {G^{\ast}}^2 \rho^{\ast}
\end {array}
\right)
\]

\vspace{0.05in}

 The values of $\omega$ are determined by vanishing of the
determinant of the above matrix. The two eigenvalues are zero and
the remaining three are determined by solution of a cubic equation
\begin{eqnarray}\label{stability_equation}
  \omega^3 + \left[(1-a+\frac{a}{\gamma}) +
\frac{24\pi\gamma^2}{a^2
      m_4^2} {G^{\ast}}^2 \rho^{\ast}\right]\omega^2 + \left[
    \frac{a(2-\gamma)}{\gamma} + \frac{6\pi (\gamma-2)(a-2)^2}{a
      m_4^2} {G^{\ast}}^2 {\phi^{\ast}}^{2} \right.  \nonumber \\
  \left. + \frac{24\pi \gamma (\gamma
      + a - \gamma a)}{a^2 m_4^2} \rho^{\ast}\right] \omega
      + \frac{6\pi (\gamma-2)}{a m_4^2} \left[(a-2)^2{\phi^{\ast}}^2 +
    4\gamma \rho^{\ast}\right] {G^{\ast}}^2 =0 \; .
\end{eqnarray}
Solutions of this equation and therefore criterion for stability of
the fixed-points depends on the value of the anomalous dimension $a$
and and other constants in the problem.

\section{Summary}

We have analyzed in this paper long time behavior of the
cosmological equations By using the Renormalization Group method,
paying special attention to the scaling solutions in two different
cases. The first case of standard FRLW cosmology  has been widely
studied and it serves us to illustrate the method we use. Although
the system in this simple case can be analyzed by means of dynamical
system technique, this analysis depends upon the choice of
normalized variables which are not easy to find. The RG method
avoids this problem: the scale invariance solutions found describe
all the scaling solutions.

We also applied RG method to the generalized cosmologies, which
imply modifications of the Friedmann equation. The RG equations give
the scaling solutions of the system and we have shown that in the
scaling regime the potential of the scalar field and the function
which describes the modifications to the Friedmann equation have a
power law dependence on their respective variables.

The RG method is a simple but powerful tool to investigate scaling
solutions for extended cosmologies. We have proved that to get
scaling solutions, during the dynamical evolution, the scalar field
potential should be either power law or exponential and the function
G, that generalizes Friedmann equation, should be a power law of the
total density. Only the fixed point given by Eq. (\ref{non_zero})
represents a scaling solution, whereas the rest of fixed points give
cosmologies with either vanishing fluid or vanishing scalar field.
The stability of the scaling solution fixed point is given by Eq.
(\ref{stability_equation}), and it depends on the particular model
under study through the constants appearing in that equation.
However, the most relevant of these constants is the anomalous
dimension $"a"$. Finally, it would be interesting to study the
evolution of a larger variety of field models as K-essence, phantom,
quintaessence, etc. (see \cite{cop-sami} for a review) with this
technique.

\section*{Acknowledgements}
We would like to acknowledge useful discussions with Takahiro Tanaka
and participants of the JGRG14, and YKIS 2005, held in Kyoto. SJ
acknowledges support under a JSPS fellowship, and JI  acknowledges
financial support under grant FIS 2004-01626.

%-------------------------------------------------------------

%----------------------------------------------------------------

\end{document}